\documentclass[twocolumn,prb,amsmath,amssymb,floatfix,superscriptaddress,showpacs]{revtex4-1}
\usepackage{graphicx}
\usepackage{dcolumn}
\usepackage{bm}
\usepackage{xspace}
\usepackage{CJK}
\begin{document}
\begin{CJK*}{GBK}{}

\title{Magnetic order tuned by Cu substitution in Fe$_{1.1-z}$Cu$_z$Te}
\author{Jinsheng Wen}
\affiliation{Department of Physics, University of California, Berkeley, California 94720, USA.}
\affiliation{Materials Science Division, Lawrence Berkeley National Laboratory, Berkeley, California 94720, USA}
\author{Zhijun Xu}
\affiliation{Condensed Matter Physics and Materials Science
Department, Brookhaven National Laboratory, Upton, New York 11973,
USA}
\author{Guangyong Xu}
\affiliation{Condensed Matter Physics and Materials Science
Department, Brookhaven National Laboratory, Upton, New York 11973,
USA}
\author{M. D. Lumsden}
\affiliation{Quantum Condensed Matter Division, Oak Ridge National Laboratory, Oak Ridge,
Tennessee 37831, USA.}
\author{P. N. Valdivia}
\affiliation{Department of Materials Science and Engineering, University of California, Berkeley, California 94720, USA.}
\author{E. Bourret-Courchesne}
\affiliation{Life Sciences Division, Lawrence Berkeley National Laboratory, Berkeley, California 94720, USA}
\author{Genda Gu}
\affiliation{Condensed Matter Physics and Materials Science
Department, Brookhaven National Laboratory, Upton, New York 11973,
USA}
\author{Dung-Hai Lee}
\affiliation{Department of Physics, University of California, Berkeley, California 94720, USA.}
\affiliation{Materials Science Division, Lawrence Berkeley National Laboratory, Berkeley, California 94720, USA}
\author{J.~M.~Tranquada}
\affiliation{Condensed Matter Physics and Materials Science Department, Brookhaven National Laboratory, Upton, New York 11973, USA}
\author{R. J. Birgeneau}
\affiliation{Department of Physics, University of California, Berkeley, California 94720, USA.}
\affiliation{Materials Science Division, Lawrence Berkeley National Laboratory, Berkeley, California 94720, USA}
\affiliation{Department of Materials Science and Engineering, University of California, Berkeley, California 94720, USA.}
\date{\today}

\begin{abstract}
We study the effects of Cu substitution in Fe$_{1.1}$Te, the non-superconducting parent compound of the iron-based superconductor, Fe$_{1+y}$Te$_{1-x}$Se$_x$, utilizing neutron scattering techniques. It is found that the structural and magnetic transitions, which occur at $\sim60$~K without Cu, are monotonically depressed with increasing Cu content. By 10\%\ Cu for Fe, the structural transition is hardly detectable, and the system becomes a spin glass below 22~K, with a slightly incommensurate ordering wave vector of (0.5-$\delta$,~0,~0.5) with $\delta$ being the incommensurability of 0.02, and correlation length of 12~\AA\ along the $a$ axis and 9~\AA{} along the $c$ axis. With 4\%\ Cu, both transition temperatures are at 41~K, though short-range incommensurate order at (0.42,~0,~0.5) is present at 60~K. With further cooling, the incommensurability decreases linearly with temperature down to 37~K, below which there is a first order transition to a long-range almost-commensurate antiferromagnetic structure. A spin anisotropy gap of 4.5~meV is also observed in this compound. Our results show that the weakly magnetic Cu has large effects on the magnetic correlations; it is suggested that this is caused by the frustration of the exchange interactions between the coupled Fe spins.
\end{abstract}


\maketitle
\end{CJK*}

\section{Introduction}
Despite the lack of a consensus on the pairing mechanism in the Fe-based superconductors on the theoretical side,~\cite{2011arXiv1106.3712H,PhysRevLett.104.157001} experimentally, there is accumulating evidence pointing to strong coupling between magnetism and superconductivity.~\cite{lumsdenreview1} It is found that superconductivity occurs in close vicinity to the magnetic instability in many systems.~\cite{lynn-2009-469} Upon chemical doping or application of pressure, the long-range antiferromagnetic order in the parent compound is suppressed, followed by the appearance of superconductivity.~\cite{dcjohnstonreview} This proximity of antiferromagnetism and superconductivity is similar to that observed in the cuprate superconductors,\cite{birgeneau-2006} suggesting that these correlations might be connected in a similar way. Moreover, a neutron-spin ``resonance" mode, where the spectral weight at a particular energy ($E_r$) and wave vector is greatly enhanced below the superconducting transition temperature, $T_c$, is widely observed,\cite{lumsdenreview1} with the resonance energy scaling roughly with $T_c$.\cite{yu-2009,qiu:067008}

In the Fe$_{1+y}$Te$_{1-x}$Se$_x$ (11) system, which is the structurally simplest system among the Fe-based superconductors,~\cite{hsu-2008} magnetism is also considered to play an important role in the superconductivity~\cite{interplaywen}; however, the situation is more complicated than in the iron pnictides. In the latter materials, the magnetic order occurs at the in-plane wave vector (0.5,~0.5) (assuming unit cell containing two Fe).  The static order is suppressed with doping, but dynamic magnetic correlations that evolve into the resonance remain centered at the same wave vector.\cite{lumsdenreview1}  In contrast, there are two characteristic wave vectors for the 11 system.~\cite{qiu:067008} The magnetic order has a characteristic wave vector around (0.5,~0) in the parent compound,~\cite{bao-2009,li-2009-79} while the resonance is observed around (0.5,~0.5) in the superconducting compound.~\cite{qiu:067008,fieldeffectresonancewen} The spectral weight at (0.5,~0) associated with the magnetic order gradually shifts to (0.5,~0.5) with Se doping.~\cite{lumsden-2009,liupi0topp} With Se substitution, the structural and magnetic phase transitions are suppressed,~\cite{spinglass} and the amount of excess Fe is reduced,~\cite{sales:094521} while bulk superconductivity gradually develops. The amount of excess Fe is an important parameter in controlling both the superconducting and magnetic properties. The excess Fe resides in an interstitial site centered above (or below) a plaquette of four Fe sites.\cite{PhysRevB.81.094115,prbliudiffuse}  The magnetic interaction between an interstitial and the Fe nearest-neighbors results in locally ferromagnetic correlations\cite{2011arXiv1103.5073Z,2011arXiv1109.5196T} and weak localization of the charge carriers,\cite{zhang:012506} and acts as a magnetic electron donor.~\cite{zhang:012506}  The excess Fe concentration is anticorrelated with the superconducting volume fraction and results in short-range magnetic order.\cite{xudoping11,2012arXiv1202.4152S}

The magnetic order in the parent compound Fe$_{1+y}$Te is also modified by excess Fe.
Specifically, the order changes as $y$ increases through the critical excess Fe content, $y_c$, from bicollinear commensurate to helical incommensurate.\cite{bao-2009,li-2009-79,PhysRevB.84.064403,2011arXiv1108.5968Z,2011arXiv1111.4236P}  The change in magnetic order correlates with the change in the low-temperature structural symmetry from monoclinic to orthorhombic.\cite{PhysRevB.84.064403,arXiv:1202.2484}  The precise value of $y_c$ depends on whether it is determined chemically from the total Fe content or by refinement of neutron diffraction data.\cite{PhysRevB.84.064403,2011arXiv1108.5968Z}  For comparison with the present work, the relevant value is $y_c\approx 0.1$.\cite{2011arXiv1108.5968Z}

The nature of the magnetic order in the 11 system has been controversial.   Early theoretical analyses of the iron pnictides suggested that the magnetism could be understood in terms of nesting between electron and hole pockets.\cite{kuroki-2008,mazin:057003}  Angle-resolved photoemission spectroscopy (ARPES) measurements~\cite{xia037002,PhysRevB.82.165113} have observed a Fermi-surface topology of the 11 system which is similar to that predicted by density functional theory (DFT) calculations~\cite{subedi-2008-78};  however, the in-plane ordering wave vector in the 11 system is incompatible with nesting.~\cite{bao-2009,structure6,xia037002} Later first-principles calculations have identified the role of local moments and the importance of Hund's exchange coupling, and have provided a more convincing description of the experimental observations.~\cite{ma-2009,fang-2009,weiguounified,2011arXiv1104.3454Y} A recent study on Fe$_{1.1}$Te has shown that the localized electrons that contribute to magnetism are not isolated from the itinerant ones, but instead are entangled, as indicated by a temperature-induced enhancement of the instantaneous magnetic moment.~\cite{2011arXiv1103.5073Z}

In this paper, we use Cu to replace Fe in Fe$_{1.1}$Te, aiming to study the impact of the weakly magnetic Cu dopants on the magnetic order, which we characterize with neutron scattering. We find that Cu drives down the structural and magnetic transitions, with long-range nearly-commensurate magnetic order retained in Fe$_{1.06}$Cu$_{0.04}$Te, but only short-range incommensurate order in FeCu$_{0.1}$Te. In the latter sample, the structural phase transition is not obvious and a transition to a spin-glass state is found at 22~K. In Fe$_{1.06}$Cu$_{0.04}$Te, the initial structural and magnetic ordering occurs at 41~K, involving short-range incommensurate order that abruptly shifts to long-range nearly-commensurate order below 36~K. Inelastic scattering measurements indicate a spin anisotropy gap of 4.5~meV in the nearly-commensurate phase.

\section{Experiment}
The single-crystal samples were grown by the Bridgman method. The raw materials (99.999\% Te, 99.98\% Fe, and 99.98\% Cu) were weighed and mixed with the desired molar ratio, and then doubly sealed into evacuated, high-purity (99.995\%) quartz tubes. The doubly-sealed materials were put into the furnace horizontally with the following sequence: ramp to 660~$^\circ$C in 3 hrs; hold for 1~hr; ramp to 900~$^\circ$C in 2~hrs; hold for 1~hr; ramp to 1000~$^\circ$C in 1~hr; hold for 12~hrs; cool to 300~$^\circ$C with a cooling rate of $-0.5$ or $-1$~$^\circ$C/hr; the furnace was then shut down and cooled to room temperature. In the furnace, there was a small temperature gradient from one end to the other (e.g., at 850~$^\circ$C, $\Delta T/$distance~$\approx5$~$^\circ$C/cm), so that the melted liquid crystallized unidirectionally. We have obtained large-size high-quality single crystals using this method.~\cite{interplaywen}

The bulk susceptibility was measured using a superconducting quantum interference device (SQUID) magnetometer. Neutron scattering experiments were carried out on the triple-axis
spectrometer HB-3 located at the High Flux Isotope Reactor, Oak Ridge National Laboratory, using
beam collimations of $48'$--$40'$--Sample--$40'$--$120'$ with a fixed final energy $E_f = 14.7$~meV. Two pyrolytic graphite filters were placed after the sample to reduce contamination from higher-order neutrons. Single crystals used in the neutron scattering experiments had a mass of 5.9~g and a sample mosaic spread of 1.8$^\circ$ for the (200) peak for Fe$_{1.06}$Cu$_{0.04}$Te, and 14.5~g with a mosaic spread of 1.5$^\circ$ for FeCu$_{0.1}$Te. The crystals were mounted in aluminum sample holders and loaded into a close-cycle refrigerator (CCR). The lattice constants were $a\approx b\simeq3.81(5)$~\AA, and $c\simeq6.25(7)$~\AA\ for Fe$_{1.06}$Cu$_{0.04}$Te, and $a\approx b\simeq3.82(6)$~\AA, and $c\simeq6.28(7)$~\AA\ for FeCu$_{0.1}$Te, using the notation where there are two Fe atoms in one unit cell. The data were collected in the $(H0L)$ scattering plane, defined by two vectors [100] and [001], and described in reciprocal lattice units (rlu) of $(a^*, b^*,c^*)=(2\pi/a,2\pi/b,2\pi/c)$.

\section{Results}
\subsection{Long-range nearly-commensurate and short-range incommensurate order in Fe$_{1.06}$Cu$_{0.04}$Te}
\begin{figure}[t]
\includegraphics[width=0.8\linewidth]{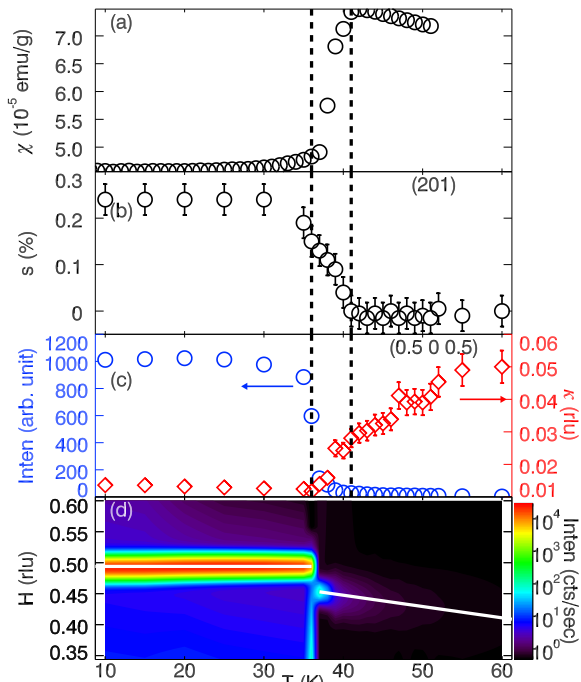}
\caption{(Color online) Magnetic and structural phase transition in Fe$_{1.06}$Cu$_{0.04}$Te. (a) Magnetic susceptibility measured with a 0.1-T filed under zero-field cooling (ZFC) condition; (b) Lattice distortion $s$ defined as $2(a-b)/(a+b)$ obtained from fitting the scans in Fig.~\ref{qscans04}(d); (c) Integrated intensity (left axis) and HWHM  (right axis) of the magnetic peak obtained by fitting the $H$ scans through (0.5,~0,~0.5) using Lorentzian functions; (d) A contour map showing a series of $H$ scans through (0.5,~0,~0.5) at different temperatures. Solid lines indicate the magnetic peak position. Vertical dashed lines indicate the transition temperatures as explained in the text.}
\label{orderp04}
\end{figure}

\begin{figure}[t]
\includegraphics[width=0.8\linewidth]{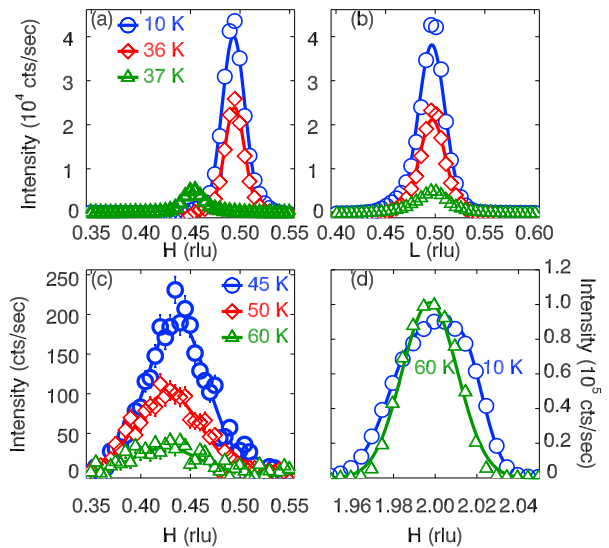}
\caption{(Color online) (a) and (b) Scans along $H$ and $L$ direction through the magnetic peak (0.5,~0,~0.5) at 10, 36, and 37~K in Fe$_{1.06}$Cu$_{0.04}$Te. (c) Scans along $H$ direction through (0.5,~0,~0.5) at 3 higher temperatures than in (a). (d) Scans along $H$ direction through the nuclear Bragg peak (201) at 10 and 60~K. Lines through data in (a)-(c) are fits using Lorentzian function. In (d), the 10-K data is fitted with a two-peak Gaussian function and the 60-K data is fitted with a single-peak Gaussian.}
\label{qscans04}
\end{figure}

The susceptibility for Fe$_{1.06}$Cu$_{0.04}$Te is plotted in Fig.~\ref{orderp04}(a). One can see that a transition to an antiferromagnetic state begins at $\sim$~41~K, as marked by the dashed line. We have performed scans through the nuclear Bragg peak (201) at various temperatures. In Fig.~\ref{qscans04}(d), we plot the results of $H$ scans through (201) at 10 and 60~K. At 60~K, the scan can be fitted with a single-peak Gaussian function reasonably well. At 10~K, although we do not see two well-resolved peaks, because of the relatively large sample mosaic and coarse resolution, we do see that the peak is significantly broader than the one at 60~K. The scan is fitted with two Gaussians. Strains $s$ defined as $2(a-b)/(a+b)$ determined from the fittings to the scans are plotted in Fig.~\ref{orderp04}(b). Assuming $s\approx0$ for $T>41$~K, an onset of non-zero strain is observed below 41~K, coincident with the drop in bulk susceptibility.  The strain grows with cooling, saturating in the range $30~{\rm K}<T<35~{\rm K}$.  The saturation correlates approximately with the onset of nearly-commensurate magnetic order, which we discuss next.

To study the magnetic order, we have performed $H$ and $L$ scans through the reciprocal lattice position (0.5,~0,~0.5) at a series of temperatures; some of the results are shown in Fig.~\ref{qscans04}(a)-(c). At low temperatures, the magnetic order is nearly commensurate, centered at (0.494,~0,~0.5). The magnetic peaks are resolution-limited along both the $H$ and $L$ directions, indicating that the order is long-ranged in three dimensions. At 37~K, the magnetic peak intensity is greatly reduced from that at 36~K, and the order becomes short-ranged and the incommensurability $\delta$ in (0.5-$\delta$,~0,~0.5) is much larger. The incommensurability increases linearly from 0.05~rlu at 37~K to 0.08~rlu at 60~K, as can be seen from Fig.~\ref{qscans04}(c).

The temperature dependence of the integrated intensity and the peak width $\kappa$  [half width at half maximum (HWHM)] of the $H$ scans through (0.5,~0,~0.5) are summarized in Fig.~\ref{orderp04}(c). At 41~K, where combined magnetic and structural transitions occur as indicated by the susceptibility and strain in Fig.~\ref{orderp04}(a) and (b), the magnetic peak intensity starts to rise more rapidly and $\kappa$ decreases. Between 37 and 36~K, there is a first-order magnetic transition, with the magnetic ordering wave vector jumping to a nearly-commensurate value. As illustrated in the contour map in Fig.~\ref{orderp04}(d), the ordering wave vector is locked below 36~K, with a small constant incommensurability of 0.006~rlu.

\begin{figure}[b]
\includegraphics[width=0.8\linewidth]{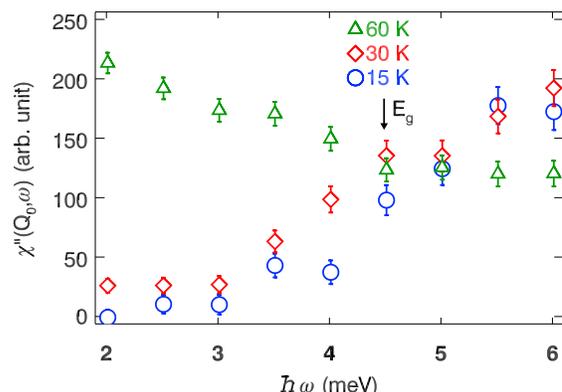}
\caption{(Color online) Energy scans from 2 to 6~meV at ${\bf Q_0}=(0.5,~0,~0.5)$ at 15 and 30~K, and at (0.42,~0,~0.5) at 60~K for Fe$_{1.06}$Cu$_{0.04}$Te. Intensity [$I$(${\bf Q_0},\omega$)] after subtracting the background obtained from the constant-energy scans has been converted to the dynamical susceptibility $\chi''({\bf Q_0},\omega)$ using $\chi''({\bf Q_0},\omega)$=$I$(${\bf Q_0},\omega$)$[1-\exp(-\hbar\omega/k_{\text B}T)]$. Arrow indicates the position of the spin anisotropy gap $E_g$.}
\label{gap04}
\end{figure}

\begin{figure}[t]
\includegraphics[width=0.8\linewidth]{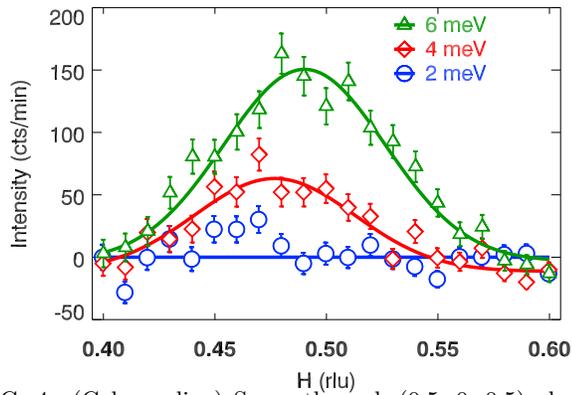}
\caption{(Color online) Scans through (0.5,~0,~0.5) along $H$ direction at 2, 4, and 6~meV at 15~K in Fe$_{1.06}$Cu$_{0.04}$Te. A sloping background has been subtracted from all data. Lines through data are fits with Gaussian function.}
\label{ineqscans04}
\end{figure}

To characterize the spin fluctuations, we have also performed inelastic scans. In Fig.~\ref{gap04}, we show energy scans at the commensurate wave vector (0.5,~0,~0.5) at low temperatures (15 and 30~K) and at the incommensurate wave vector (0.42,~0,~0.5) at a high temperature (60~K). From Fig.~\ref{gap04}, we identify a spin anisotropy gap $E_g$ of $\sim$~4.5~meV. Below $E_g$, the spectral weight is greatly suppressed at low temperature. The suppression is more clearly demonstrated in Fig.~\ref{ineqscans04}, where scans through (0.5,~0,~0.5) along the $H$ direction at 2, 4, and 6~meV are plotted for $T=15$~K. At 6~meV, above $E_g$, there is a peak centered near the commensurate wave vector. At 2~meV, well below $E_g$, the spectral weight is almost completely suppressed. This behavior is similar to that in the Cu-free materials Fe$_{1+y}$Te,~\cite{2011arXiv1103.1811,2011arXiv1111.4236P} except that the gap value is larger (7~meV) in Fe$_{1.057}$Te.~\cite{2011arXiv1103.1811}

\subsection{Short-range incommensurate order in FeCu$_{0.1}$Te}

\begin{figure}[t]
\includegraphics[width=0.8\linewidth]{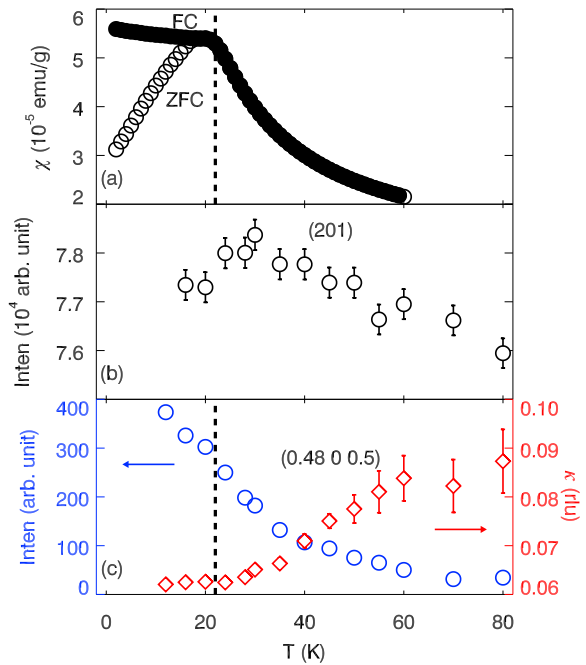}
\caption{(Color online) Short-range incommensurate magnetic order in FeCu$_{0.1}$Te. (a) Susceptibility measured with 0.1-T filed under ZFC and FC conditions; (b) Intensity vs. temperature of the (201) Bragg peak; (c) Integrated intensity (left axis) and HWHM (right axis) obtained by fitting the $H$ scans through the magnetic peak (0.48,~0,~0.5) with Lorentzians. Dashed lines indicate the spin-glass transition temperature.}
\label{orderp10}
\end{figure}

Now we turn to the 10\% Cu-substituted sample, FeCu$_{0.1}$Te. The magnetic properties of this sample are somewhat different from those of Fe$_{1.06}$Cu$_{0.04}$Te. The susceptibility data are shown in Fig.~\ref{orderp10}(a). At $\sim$~22~K, there is a cusp in the susceptibility, with hysteresis between field-cooling (FC) and zero-field cooling (ZFC) at lower temperatures. The susceptibility data look similar to those reported in Ref.~\onlinecite{spinglass} for Fe$_{1+y}$Te$_{1-x}$Se$_x$ with $x=$ 15\% and 30\%, both of which are spin-glass-like. We determine that our sample is also a spin glass at low temperature, which is expected to have no long-range order. The spin-glass transition temperature is marked by the dashed line. We have searched for a signature of lattice distortions in this sample. Throughout the whole temperature range, the shape of the nuclear Bragg peak (201) remains unchanged within our limited resolution. The peak intensity, as plotted in Fig.~\ref{orderp10}(b), does show a maximum around 30~K, which might correspond to a slight broadening at lower temperature. Any lattice distortion is much less than that observed for Fe$_{1.06}$Cu$_{0.04}$Te.

\begin{figure}[b]
\includegraphics[width=0.8\linewidth,trim=-7mm 75mm -3mm -1mm,clip]{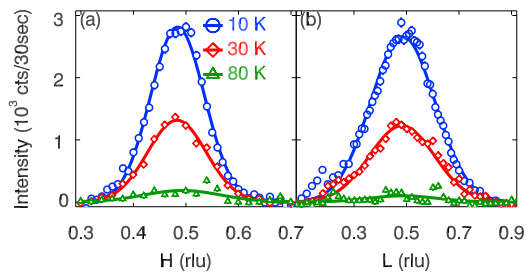} 
\caption{(Color online) Scans along $H$ and $L$ direction through the magnetic peak (0.5,~0,~0.5) in FeCu$_{0.1}$Te.}
\label{qscans10}
\end{figure}

Scans along the $H$ and $L$ directions through the magnetic peak (0.5,~0,~0.5) at 3 representative temperatures are shown in Fig.~\ref{qscans10}. There is a weak, broad peak already at 80~K, which gradually begins to narrow and intensify on cooling below 60~K. Lorentzian fits yield a peak position of (0.48,~0,~0.5), corresponding to a small incommensurability of $0.02$~rlu along the $H$ direction.  The integrated intensity and peak width (HWHM) $\kappa$ obtained from fits to $H$ scans are plotted as a function of temperature in Fig.~\ref{orderp10}(c). The intensity increases from 60~K through $T_{sg}$. Below $T_{sg}$, $\kappa$ saturates at $\sim$~0.06~rlu. By taking into account the resolution of $\sim$~0.01~rlu, this corresponds to a correlation length of 12~\AA{} along the $a$ axis. The order along the $c$ axis is also short-ranged, with a correlation length of 9~\AA{}. These values are comparable with those in Fe$_{1+y}$Te$_{1-x}$Se$_x$ with $x\agt0.2$.~\cite{wen:104506,li-2009-79}

\section{Discussions and Conclusions}
\subsection{Tuning of the magnetic order}

In Fe$_{1+y}$Te, experiments have established that the ground state crystal and magnetic structures change with $y$.\cite{bao-2009,PhysRevB.84.064403,arXiv:1202.2484}   For $y<y_c\approx 0.1$, the structure is monoclinic, with two different Fe-Fe in-plane bond lengths that match up with the alternating antiferromagnetic and ferromagnetic bonds of the collinear magnetic structure.\cite{bao-2009,PhysRevB.81.094115,paul11} For $y>y_c$, the structure is orthorhombic, with a single Fe-Fe bond length;  the corresponding magnetic structure is found to be helical.\cite{bao-2009,PhysRevB.84.064403} Mizuguchi {\it et al.}\cite{arXiv:1202.2484} have recently demonstrated with high-resolution x-ray diffraction that the boundary between the orthorhombic and monoclinic phases can also be crossed as a function of temperature for $y\lesssim y_c$.  On cooling from the tetragonal phase, it is possible to transform first to the orthorhombic and then to the monoclinic.  This series of transitions occurs in the same regime where the initial magnetic order at high temperature is an incommensurate spin density wave, that transforms abruptly to a nearly-commensurate version of the bicollinear structure.\cite{PhysRevB.84.064403,2011arXiv1108.5968Z}

The bicollinear magnetic structure can be obtained from DFT calculations\cite{ma-2009,johannes-2009,moon057003}; however, interpretation of the components determining the magnetic structure are usually done in terms of simpler model Hamiltonians.\cite{fang-2009,loalm1,weiguounified}  It is clear that because of competing interactions, the bicollinear and collinear antiferromagnetic states\cite{lynn-2009-469} are close in energy.  While the magnetic structures can be rationalized with a Heisenberg model\cite{fang-2009} and biquadratic spin terms are expected from spin-lattice coupling,\cite{loalm1} itinerant electrons are also expected to play a role.\cite{weiguounified}  Yin {\it et al.}\cite{weiguounified} have argued that the itinerant component, combined with strong Hund's coupling, leads to double-exchange interactions that favor ferromagnetic bonds, similar to the case of manganites.  This, in turn, can result in a local breaking of the degeneracy of the Fe $d_{xz}$ and $d_{yz}$ orbitals.  It seems possible that such spin-orbit coupling effects may contribute to the spin anisotropy gap that is experimentally observed to be large in the bicollinear state.\cite{2011arXiv1103.1811}

A feature that is more difficult to include in the models is the interstitial Fe.  In a study of Fe$_{1.1}$Te,\cite{2011arXiv1103.5073Z} broad incommensurate diffuse scattering was observed that could be well described by a model of local ferromagnetic plaquettes of Fe sites, with antiferromagnetic correlations between plaquettes.  Such correlations are close in energy to the bicollinear state,\cite{2011arXiv1106.0881Y} and a ferromagnetic plaquette provides the best coupling to a magnetic interstitial.  It seems likely that such interactions play a role in the incommensurate order observed above the transition to the bicollinear state in Fe$_{1+y}$Te with $y\approx0.1$.\cite{PhysRevB.84.064403,2011arXiv1108.5968Z}  Parshall {\it et al.}\cite{2011arXiv1111.4236P} have discussed evidence for a competition between commensurate and incommensurate order.

The transitions in our Fe$_{1.06}$Cu$_{0.04}$Te sample appear quite similar to those reported for Fe$_{1+y}$Te with $y\approx0.1$,\cite{PhysRevB.84.064403,2011arXiv1108.5968Z} but with reduced ordering temperatures.  This observation suggests that the Cu substitutes for in-plane Fe, causing a reduction in ordering due to quenched disorder.  The small moment of a Cu ion and shifted $d$ levels will frustrate both double-exchange and superexchange interactions.  The abrupt onset of the bicollinear order at 36~K presumably coincides with an unresolved transition to monoclinic order.  It seems likely that the structure between 36 and 41 K is orthorhombic, based on the patterns discussed above.  The reduction of the anisotropy gap to 4.5 meV, compared to 7 meV in a sample without Cu,\cite{2011arXiv1103.1811} is consistent with reduction in the average interactions due to magnetic dilution and/or frustration.

For FeCu$_{0.1}$Te, the quenched disorder is sufficiently disruptive that the structural transition appears to be washed away and the low-temperature magnetic order is reduced to a cluster-glass state.  The magnetic ordering is similar to a low-temperature extrapolation of the incommensurate phase  of the Fe$_{1.06}$Cu$_{0.04}$Te sample.  The incommensurability and short correlation lengths are also similar to those observed in Fe$_{1+y}$Te$_{1-x}$Se$_x$ with $x\agt0.2$.\cite{wen:104506,2011arXiv1109.5196T,spinglass}

\subsection{Relationship to superconductivity}

\begin{figure}[b]
\includegraphics[width=0.8\linewidth]{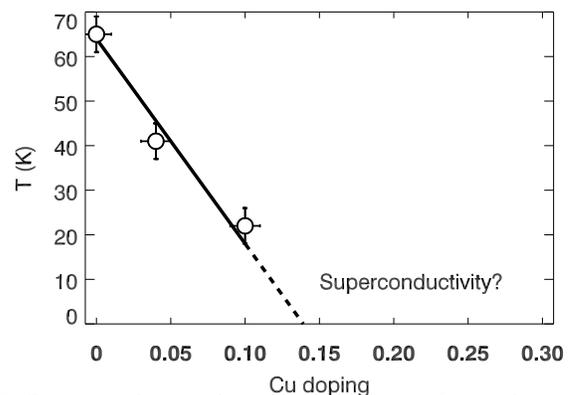}
\caption{Magnetic ordering temperature vs. Cu doping. Lines are guides to the eye.}
\label{phasedgm}
\end{figure}

In many of the Cu- and Fe-based superconductors, superconductivity develops concomitantly with the
suppression of the structural phase transition and antiferromagnetic order.~\cite{birgeneau-2006,dcjohnstonreview} This also holds for Fe$_{1+y}$Te$_{1-x}$Se$_x$. When Se is used to replace Te, the lattice distortion and antiferromagnetic order are suppressed, followed by the appearance of bulk superconductivity with $x\agt0.3$.~\cite{spinglass} Substitution of Te with Se is believed to frustrate the antiferromagnetic order.~\cite{PhysRevB.84.064403,fang-2009} When we substitute Fe with Cu, the trend appears to be similar, as shown in Fig.~\ref{phasedgm}. This naturally raises the question of whether or not superconductivity can be induced with further Cu doping. Our initial attempts have indicated no signature of superconductivity with Cu substitution up to 0.3.~\cite{wenunpublished}

It appears that the suppression of the lattice distortions and antiferromagnetic order is not sufficient to produce superconductivity.  Similar observations have been made in related systems. In a recent report, the isoelectronic substitution of Sb for As in LaFeAsO  reduces the structural and magnetic transition temperatures, but it does not induce superconductivity.~\cite{PhysRevB.84.104523} An NMR study on LaFe$_{1-x}$Ru$_x$AsO sees the dilution of antiferromagnetism with Ru doping, but without the emergence of superconductivity.~\cite{PhysRevB.85.054518} The electronic and structural properties, such as the anion height from the Fe layer, and the bonding angle between the anion and the Fe may also need to be optimized in order to have robust superconductivity.~\cite{PhysRevB.79.224511,JPSJ.79.102001} Furthermore, adding Cu into Fe$_{1+y}$Te$_{1-x}$Se$_x$ is believed to introduce a localized state,~\cite{0953-8984-22-13-135501} which is deleterious for superconductivity.~\cite{williams-2009-21,PhysRevB.82.104502} Of course, we could be simply blocked out from accessing the portion of the phase diagram by reaching the solubility limit. As in the case of Fe$_{1+y}$Se, the maximum amount of Cu that can be substituted into the compound is 20--30\%.~\cite{williams-2009-21}

The key issue is likely associated with the magnetic wave vector.  The presence of interstitial Fe favors magnetic correlations near $(0.5,~0)$, as already discussed.  On the other hand, a clear connection between magnetic correlations and superconductivity is only observed when the low-energy fluctuations occur near $(0.5,~0.5)$.\cite{lumsdenreview1}  Interstitial Fe in Se-doped samples is known to disrupt the superconductivity and to shift the magnetic wave vector.\cite{2012arXiv1202.4152S}

\subsection{Conclusions}
In summary, we have shown that Cu substitution of Fe in Fe$_{1.1-z}$Cu$_z$Te has a similar effect to that of Se substitution of Te, as they both suppress the lattice distortion and magnetic order through the impact of quenched disorder. In a localized model, this implies that with increasing Cu/Se, the average exchange couplings between the Fe spins are reduced, or the frustration increases, leading to the suppression of the magnetic order. However, this behavior is different from the impact of interstitial Fe, which results in different types of magnetic ground states rather than destroying the order.

\section{Acknowledgements}
The work at Lawrence Berkeley National Laboratory and Brookhaven National Laboratory was supported by the Office of Basic Energy Sciences, Division of Materials Science and Engineering, U.S. Department of Energy, under Contract No.\ DE-AC02-05CH11231 and DE-AC02-98CH10886 respectively. Research at Oak Ridge National Laboratory's High Flux Isotope Reactor was sponsored by the Division of Scientific User Facilities of the same Office.

%

\end{document}